\documentclass[12pt,a4paper]{article}

\usepackage[textheight=23cm,textwidth=15cm]{geometry}
\usepackage{amsmath}
\usepackage{graphicx}
\usepackage[american]{babel}
\usepackage{here}
\usepackage[articletitle=true]{achemso}
\usepackage{url}

\begin{document}
\begin{center}
{\LARGE\bf
High-throughput ab initio reaction mechanism exploration in the cloud with automated multi-reference validation
}

\vspace{0.5cm}

{\large
Jan P. Unsleber$^{a,}$\footnote{ORCID: 0000-0003-3465-5788},
Hongbin Liu$^{b,}$\footnote{Corresponding author; e-mail: hongbin.liu@microsoft.com; ORCID: 0000-0001-9011-1182},
Leopold Talirz$^{c,}$\footnote{ORCID: 0000-0002-1524-5903},
Thomas Weymuth$^{a,}$\footnote{ORCID: 0000-0001-7102-7022},
Maximilian M\"orchen$^{a,}$\footnote{ORCID: 0000-0002-7467-5719},
Adam Grofe$^{b,}$\footnote{ORCID: 0000-0002-8531-4396},
Dave Wecker$^{b}$,
Christopher J. Stein$^{d,}$\footnote{ORCID: 0000-0003-2050-4866},
Ajay Panyala$^{e,}$\footnote{ORCID: 0000-0002-0846-3347},
Bo Peng$^{e,}$\footnote{ORCID: 0000-0002-4226-7294},
Karol Kowalski$^{e,}$\footnote{ORCID: 0000-0001-6357-785X},
Matthias Troyer$^{b,}$\footnote{ORCID: 0000-0002-1469-9444},
Markus Reiher$^{a,}$\footnote{Corresponding author; e-mail: markus.reiher@phys.chem.ethz.ch; ORCID: 0000-0002-9508-1565}}
\\[3ex]

$^{a,}$Laboratory of Physical Chemistry and NCCR Catalysis, ETH Zurich, Vladimir-Prelog-Weg 2, 8093 Zurich, Switzerland
\\
$^{b}$ Microsoft Quantum, Redmond, WA, 98052, USA
\\
$^{c}$ Microsoft Quantum, Zurich, Switzerland
\\
$^{d}$ Technical University of Munich, TUM School of Natural Sciences, Department of Chemistry, Lichtenbergstr. 4, D-85748 Garching, Germany
\\
$^{e}$ Physical and Computational Sciences Directorate, Pacific Northwest National Laboratory, Richland, Washington 99354, USA
\\[3ex]

February 16, 2023
\end{center}

\newpage

\begin{center}
{\bf Abstract}
\end{center}
Quantum chemical calculations on atomistic systems have evolved
into a standard approach to study molecular matter. These calculations
often involve a significant amount of manual input and expertise although
most of this effort could be automated, which would alleviate the need for expertise in software and
hardware accessibility. Here, we present the AutoRXN workflow, an automated workflow
for exploratory high-throughput electronic structure calculations of
molecular systems, in which (i) density functional theory methods 
are exploited to deliver minimum and transition-state structures and
corresponding energies and properties, (ii) coupled cluster 
calculations are then launched for optimized structures to provide 
more accurate energy and property estimates, and (iii) multi-reference
diagnostics are evaluated to back check the coupled cluster results and subject them to automated multi-configurational calculations for potential multi-configurational cases. All calculations are carried out in a cloud environment and support massive computational campaigns. Key features of all components of the AutoRXN workflow are autonomy, stability, and minimum operator interference.  
We highlight the AutoRXN workflow with the example of an autonomous reaction
mechanism exploration of the mode of action of a homogeneous catalyst for the asymmetric reduction of ketones.

\section{Introduction}\label{introduction}
Quantum chemical calculations have become a reliable peer to experimental studies
dedicated to the elucidation of chemical mechanisms and molecular properties\cite{Dykstra05}.
Well established protocols exist that allow one to explore chemical processes with efficient quantum mechanical methods\cite{Jensen2017, Cramer2004}.
In recent years, automated procedures have been developed 
\cite{Sameera2016,Dewyer2017,Simm2019a,Maeda2021,Baiardi2022}
that enable massive computational campaigns far beyond the manual assessment
of molecular behavior. As a consequence, quantum chemical
approaches can now be used to explore uncharted territory in chemical
(reaction) space on a truly broad scale. 

The last decade has seen such campaigns first in the context of
searching for specific molecular properties in materials science; the materials genome initiative\cite{Jain2013} and the Harvard clean energy project\cite{Hachmann2011, Hachmann2014}
are prominent examples of such achievements in 
high-throughput virtual screening\cite{Pyzer-Knapp2015, Nandy2021}.
Whereas virtual screening of physical properties simply requires
a molecular equilibrium structure that is usually straightforward to obtain,
chemical reaction kinetics require the inspection of relevant
paths across Born--Oppenheimer potential energy surfaces, which is
a far more complex task compared to molecular structure optimization.
Still, even for this purpose automated procedures have been developed
(\textit{cf.}, Refs.~\citenum{Unsleber2020,Steiner2022} and references cited therein).

All of these advances benefit from easy access to suitable high-performance computing (HPC) hardware. Cloud services have emerged as
an increasingly important alternative to local HPC clusters. \cite{seritan2020,kutzner2022}
The costs for cloud access
can be much less than purchasing and maintaining own hardware nodes
in a local cluster. Especially for non-experts and non-regular need
for such computing resources a cloud solution is attractive. However, this then requires computational workflows designed for cloud systems.

In this work, we present the AutoRXN workflow, a new, largely automated workflow that allows
for easy access to quantum chemical data on molecular systems
through open-source software in a cloud computing environment.
In Sec.~\ref{cloudoverview}, we introduce the components of our workflow,
which will then be demonstrated in a case study. This case study is
the elucidation of the reaction mechanism of a homogeneous catalyst
that has served us for the tailored development of the AutoRXN workflow.

We demonstrate key advantages of the AutoRXN workflow, namely 
(i) carrying out a huge number of comparatively cheap quantum chemical calculations
for explorative purposes,
(ii) refinement of the results obtained by a vast number of expensive correlated
\textit{ab initio} calculations, and
(iii) automated collection and evaluation of data including 
back checking of results by alternative \textit{ab initio} approaches.

This work is organized as follows: in Sec.~\ref{cloudoverview}, we introduce the conceptual organization of our AutoRXN workflow.
Then, we explain the computational methodology (the prototypical catalysis application example is introduced in section \ref{chemicalexample}). Finally, section \ref{results} presents
the results obtained for our catalysis example.

\section{Conceptual Overview of the Cloud-Based High-Throughput Workflow}\label{cloudoverview}
\begin{figure}
    \centering
    \includegraphics[width=0.99\textwidth]{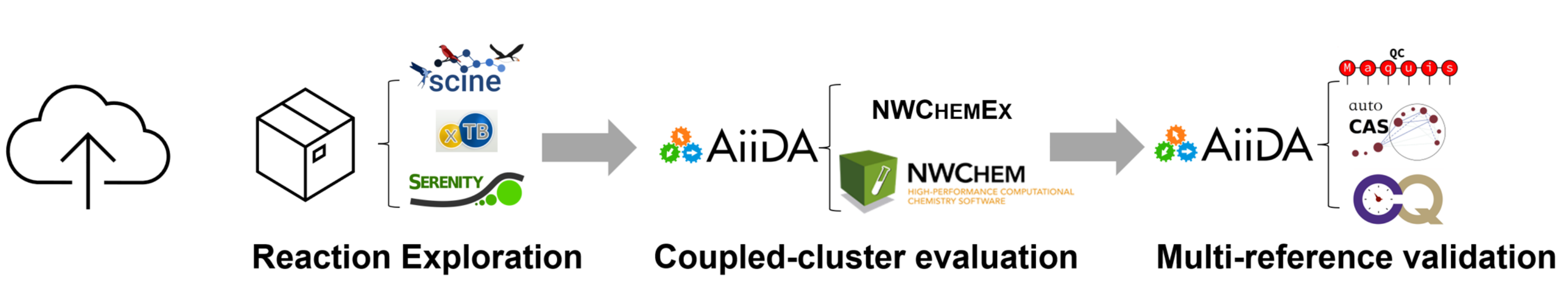}
    \caption{The overall AutoRXN autonomous reaction exploration and validation workflow developed in this work and deployed on the Azure cloud service. For a classification of the software mentioned in this figure, see text.}
    \label{fig:cloud_architecture}
\end{figure}

In this section, we introduce a conceptual overview on the AutoRXN workflow. We based our hard- and software backbone on Microsoft's cloud computing platform Azure\cite{azure}. In particular, Azure Virtual Machines (VMs)  and Azure CycleCloud have been extensively used as the computing entry-point and high-performance computing environment.\cite{azurevm,azurecyclecloud} In this environment, we implemented the workflow shown in Figure~\ref{fig:cloud_architecture} that connects three main steps of the AutoRXN workflow: 
1) A meta program that orchestrates the high-throughput screening (in our case, the \textsc{SCINE Chemoton}\cite{Moritz2022,unsleber2022} software for first-principles reaction network explorations)
with raw data generation by quantum chemistry program packages (such as the open-source \textsc{Serenity} software)\cite{Barton2021}.
2) An AiiDA\cite{pizzi2016} workflow\cite{aiida-nwchemex} that executes high-throughput \textit{ab initio} calculations, such as coupled-cluster calculations, to evaluate accurately reaction barriers and reaction energies.\cite{valiev2010,kowalski2021}
3) An AiiDA workflow\cite{aiida-autocas} that executes automated multi-reference calculations (based on fully automated active space selection \cite{Stein2016,Stein2019}) to validate coupled-cluster barriers and energies in the case of a potential multi-configurational reference detection with multi-configurational perturbation theory (implemented in QCMaquis\cite{Keller2015} in concert with ChronusQ \cite{Li2019}). 

In Sec.~\ref{methodology}, we will illustrate our strategy of implementation and explain why steps such as coupled-cluster and multi-reference calculations are implemented as standalone workflows rather than monolithic calculations integrated in the meta program.

Certain key benefits are offered by implementing the AutoRXN workflow on a cloud, and in particular on Azure: Cloud computing in general provides an optimal scale of computing, which is critical for any high-throughput task: when millions of reaction pathways need to be evaluated, idle computing resources in the geographically distributed data centers can be instantly allocated to fulfill the demand. In addition, cloud typically has zero up-front cost, which makes it more accessible to researchers without dedicated high-performance computing resources. Moreover, Azure features a very rich selection of computer hardware\cite{azurevmsize}, especially targeting HPC workloads, so that various types of calculations, from semi-empirical to multi-reference ones, can be mapped to most suitable hardware architectures for optimal acceleration.

\section{Computational Methodology}\label{methodology}

\subsection{The chemical example: asymmetric reduction of ketones} \label{chemicalexample}
We implemented and developed the AutoRXN workflow at a
transition metal catalysis example that can be considered prototypical for complex molecular transformation networks. Whereas we obtained detailed information and new insights for this 
specific system and its reactivity, our main goal is to demonstrate
the feasibility of our approach at a system of typical size. We emphasize that, by
virtue of the first-principles nature of all procedures involved,
our approach is general and can be adopted to various types of problems in 
reaction chemistry and catalysis. In other words,
our results may be taken as a clear indication of the level of detail that may
be reached for any other reactive molecular chemical system of interest.

Our specific choice here is a homogeneous iron catalyst system for
the asymmetric transfer hydrogenation of ketones
that was first reported by Morris \textit{et al.}\cite{Mikhailine2009}
This family of catalysts has been improved several times, culminating in the development of the ``third generation'' in 2013\cite{Zuo2013}. It  belongs to the most efficient iron-based catalysts for asymmetric transfer hydrogenation\cite{Seo2019}.
A first proposition of a mechanism, supported by experimental findings, was provided in Ref.~\citenum{Mikhailine2012}.

\begin{figure}
    \centering
    \includegraphics{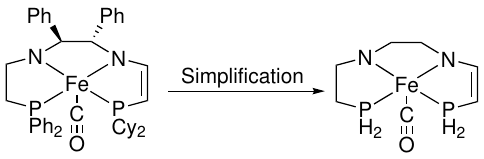}
    \caption{Original catalyst (left) reported in Ref.~\citenum{Prokopchuk2012} and simplified (generic) catalyst structure considered in this work (right).}
    \label{fig:catalyst_reduction}
\end{figure}

To reduce the complexity of the catalyst for the sake of computational efficiency in the context of the workflow development, we replaced
bulky substituents in the ligand sphere with hydrogen atoms.
While this surely compromises the
asymmetry of the chemical transformation, it is of little
effect to the workflow development process (however, care must
be taken when directly comparing our calculated results with experimental findings). Moreover, the
structural simplifications will produce a reaction network of
the generic catalyst that may be subsequently subjected to different
substitution strategies in search of optimized catalyst structures
that fulfill predefined purposes. Such structural design work may
well start from the generic reaction network and exploit the 
high-throughput workflow presented here.
The simplified catalyst structure is shown in Figure~\ref{fig:catalyst_reduction}.

A first density functional theory (DFT) investigation (with implicit solvation) into the reaction mechanism of the same simplified system as the one considered in
our work here was consistent with experimental findings\cite{Prokopchuk2012}.
A subsequent DFT study then considered the full catalyst structure\cite{Zuo2016}. 
We note in passing that the same catalyst can be utilized for asymmetric hydrogenation\cite{Zuo2014, Zuo2016}, which indicates the importance of side reactions. Hence, it is decisive for a complete understanding of such a system to also cover side and decomposition reactions.

\begin{figure}
    \centering
    \includegraphics{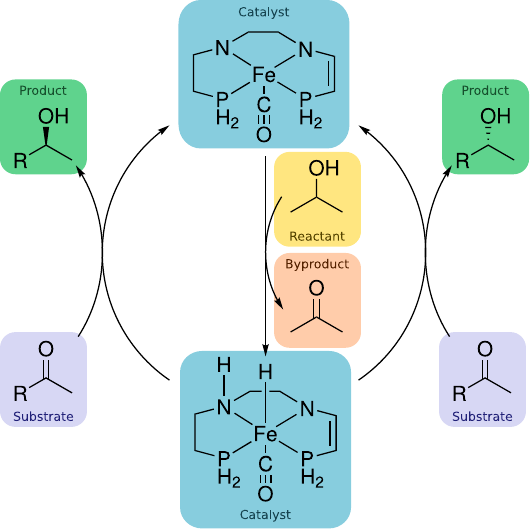}
    \caption{Overview of the catalytic process including both stereochemistries in a minimal cycle.}
    \label{fig:ath_cycle_lewis_minimal}
\end{figure}

\begin{figure}
    \centering
    \includegraphics[scale=0.9]{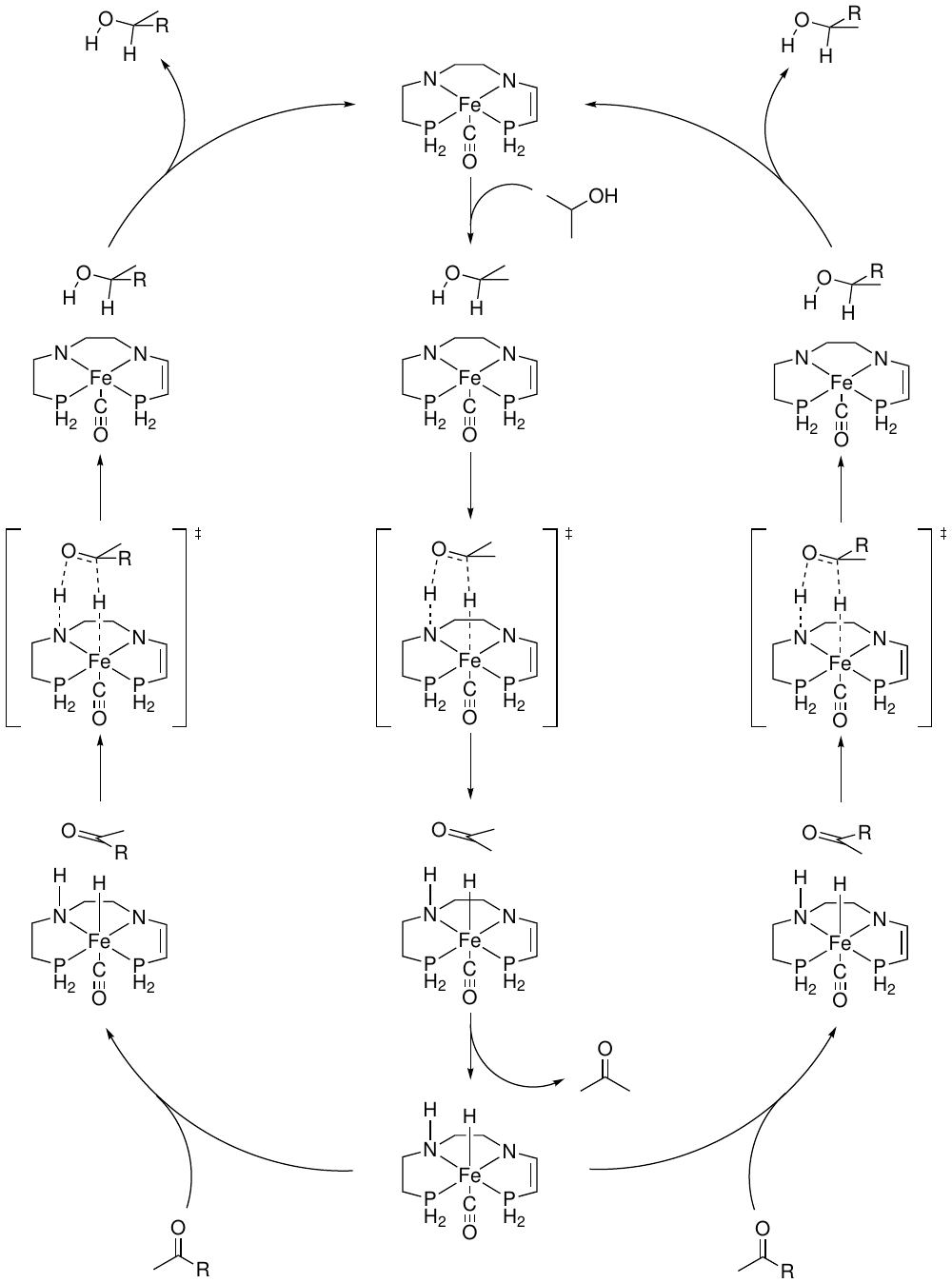}
    \caption{Expanded view of the cycle with additional structural information for association complexes and transition states.}
    \label{fig:ath_cycle_lewis}
\end{figure}

In this work, we chose to investigate the transformation of acetophenone to 1-phenylethanol.
The transferred formal dihydrogen moiety is to be provided by oxidation of isopropyl alcohol to acetone;
the minimal catalytic cycle is shown in Figure~\ref{fig:ath_cycle_lewis_minimal} to provide an overview on
the whole process.
A more extensive version, including Lewis structures of association complexes and key transition states and intermediates for both enantiomers, \textit{i.e.}, \textit{(R)} and \textit{(S)} products, is provided in Figure~\ref{fig:ath_cycle_lewis}.

In the following subsections, we describe the individual computational components of
the AutoRXN workflow as applied to the asymmetric hydrogenation catalyst. We emphasize that the development of the workflow 
required activating different components separately until the complete
pipeline was established. For instance, while the complete workflow
sketched in section \ref{cloudoverview} above
can now be activated, the efficient development of its components
was accelerated by separating individual steps (such as the molecular
structure generation).

\subsection{Molecular structure generation}\label{moleculegeneration}
The reaction network was initialized by two catalyst structures, the H$_2$-loaded catalyst and the empty catalyst (see Figure~\ref{fig:ath_cycle_lewis_minimal}). Additionally, isopropanol and acetophenone were added as nodes to the reaction network.
Based on these four starting compounds, possible elementary steps were explored with development versions of \textsc{SCINE Chemoton}\cite{Moritz2022,unsleber2022} (\textit{cf.}, the step ``Reaction Exploration'' in Figure~\ref{fig:cloud_architecture}).

The raw data for this exploration were obtained with \textsc{Turbomole} 7.4.1\cite{turbomole, Ahlrichs1989} in its shared-memory parallelized version as the backend program. DFT calculations with the PBE exchange--correlation functional\cite{Perdew1996, Perdew1997} and Ahlrichs' def2-SVP basis set\cite{Weigend2005} for all atoms were supplemented by semi-classical D3 dispersion corrections\cite{Grimme2010} (with the Becke--Johnson damping function\cite{Grimme2011}). The resolution-of-the-identity density fitting approximation was invoked, alongside with the corresponding auxiliary basis sets\cite{Weigend2006}. All calculations were carried out for isolated molecules (\textit{i.e.}, modeling the gas phase) without point group symmetry restrictions, \textit{i.e.}, in C\textsubscript{1} point group symmetry. We note, however, that a dielectric continuum model can
be easily turned on.

Initially, only electronic energies are considered in the exploration process. Thermal corrections for energies of all structures can, however, be automatically generated by \textsc{Chemoton} if requested. Here, we will show the Gibbs free energy pathways (obtained with PBE-D3BJ/def2-SVP for the harmonic frequencies in the standard particle-in-the-box harmonic-oscillator rigid-rotor approximation) for the catalytic cycle in section \ref{maincycle}. 
Second-order derivatives of the electronic energy with respect to
the nuclear Cartesian coordinates, \textit{i.e.}, nuclear Hessians, were calculated for transition states in order to validate them in an automated fashion.

For the purpose of this work, namely the development of the AutoRXN cloud-based high-throughput quantum chemistry workflow, we did not aim at ultimate depth of the autonomous exploration and therefore
restricted the exploration as follows:
The elementary step trial generation in \textsc{Chemoton} was configured to allow for bimolecular reaction trials with the Newton Trajectory 1 (NT1) type algorithm only\cite{unsleber2022}.
For all probed (potentially reactive) pairs of compounds, the 
automated exploration was restricted to only one structure (\textit{i.e.}, conformer) per compound (note that this is not a particularly severe restriction because all compounds of the catalytic cycle, \textit{cf.} Figure~\ref{fig:ath_cycle_lewis}, are well represented by a single conformer only due to their structural rigidity).
Multiple attack points per atom or reactive fragment were allowed for each of these structures, and two rotamers per trial complex were generated.
Multi-atom fragments within the NT1 algorithm were limited to atoms that are at maximum three bonds apart.
The initial compounds were filtered in such a way that only combinations of one substrate and one structure of the catalyst were tried for elementary steps, combinations of two substrates or two structures of the catalyst with one another were omitted.
No newly generated reaction products were allowed to become reactants. We emphasize that all these constraints were defined upon
input to the automated \textsc{Chemoton} exploration.

Besides this starting configuration for the automated generation of elementary step trials, the searches for elementary steps that were expected to generate the catalytic cycle proposed in the literature were prioritized over the others.
To this end, all trials intended to scan bond changes including the iron or nitrogen atoms in the catalyst structures and the oxygen and attached carbon atoms in one of the substrates were prioritized through user intervention with the exploration setup.
No explicit search for conformers was conducted, but conformational changes occurring during searches for new elementary steps were tracked.

A condensed version of the generated data is available on Zenodo\cite{dataset}.

\subsection{High-throughput coupled cluster calculations}\label{htcc}

The AutoRXN workflow allows us to carry out extensive system-focused benchmarking calculations to challenge the DFT results and to provide an alternative access to \textit{ab initio} reference data.
Since coupled-cluster singles and doubles with perturbative triples (CCSD(T)\cite{raghavachari1989}) and a sufficiently large one-particle basis set is currently regarded as the gold standard in quantum chemistry,\cite{thomas1993,helgaker1997}
the AutoRXN workflow activates this specific electronic structure model to generate accurate reference data. Since we did not apply selective techniques\cite{Simm2018} 
to transfer information on the error compared to DFT, we obtained CCSD(T) data for {\it all} structures in our reaction network. We used the \textsc{NWChemEx} Coupled Cluster software as the main engine.\cite{kowalski2021} A CCSD(T) calculation was decomposed into the following four steps: 1) Hartree--Fock (HF) calculation, 2) Cholesky decomposition (CD) of the electron repulsion integrals, 3) CD-CCSD calculation, and 4) perturbative triples correction evaluation.
The computational characteristics of the four steps are significantly different and can therefore be mapped onto different hardware for execution.

The underlying algorithms and linear algebra for HF and CD have been optimized for central processing unit (CPU) executions and do not benefit much from graphics processing unit (GPU) acceleration. 
Therefore, those two steps are executed on the  HC44rs HPC VMs on
Azure.\cite{hcvm} The CD-CCSD and CCSD(T) calculations are both GPU-accelerated in \textsc{NWChemEx}. However, the implementation of CD-CCSD requires large memory and the communication between GPUs is quite intense. Therefore, CD-CCSD is conducted on the  high-memory, Infiniband-interconnect ND40rsv2 GPU VMs on Azure. \cite{ndv2vm} By contrast, the CCSD(T) implementation features minimal communication overhead and can therefore be conducted on the low-memory, Ethernet-interconnect NC24 GPU VMs.\cite{ncvm} The streamlined execution of the four steps on heterogeneous hardware in Azure is realized through AiiDA workflows\cite{pizzi2016,aiida-nwchemex}. 

For the iron-based catalyst system studied in this paper, which is a relatively small chemical system (the number of basis functions does not exceed 500 in a cc-pVDZ \cite{balabanov2005} basis set), the runtime of each individual step is relatively short (even for CD-CCSD or CCSD(T), a single-point calculation finishes within 2 hours). Therefore, the low-priority VMs in Azure \cite{spotvm} can be fully exploited in the high-throughput calculations to reduce significantly the financial costs of the exploration.

An average-size iron-catalyzed asymmetric hydrogenation system (400 basis functions) requires $\$112$ when running CCSD(T) on the infiniband-enabled HC44rs VMs using \textsc{NwChem}\cite{valiev2010} on Azure. It only costs $\$10$ using the AiiDA-NWChemEx workflow running on low-priority VMs with a 30-fold runtime speedup.

\subsection{Driving multi-configurational calculations with \textsc{autoCAS}}\label{autocas}

Catalytic reactions, especially those involving $3d$-metal catalysts, often show strong electron correlation effects even in the ground state, but, in particular, in transition state structures due to bond formation and breaking.
In such situations, static electron correlation breaks the accuracy of single-reference methods such as coupled-cluster, because a multi-configurational wave function will be required as reference.
For an accurate description of these cases, multi-reference methods must therefore be employed.
In practice, these methods scale unfavorably with the number of correlated orbitals in a given system so that not all orbitals can be considered.
As a result, a subset of orbitals (the so-called active orbital space) must be chosen so that all orbitals responsible for static correlation effects are included.
In the past, this subset was selected manually, so that this flavor of electronic structure methods is not suited for a fully automatic exploration.

We resolved this issue with the \textsc{SCINE autoCAS} program\cite{Stein2016, Stein2019} for automatically selecting the active space based on single-orbital entanglement entropies. We note that other very recent efforts have also targeted high-throughput multi-configurational calculations.\cite{king2022}
For the automated launch of multi-reference calculations in our work here, we developed a command-line version of \textsc{autoCAS} 2.0.0\cite{Morchen2022} (\textit{cf.}, https://scine.ethz.ch/download/autocas) as a Python3 module and integrated it into the
AutoRXN workflow in such a way that it is started automatically once a T1 diagnostics is detected to be larger than 0.1 for one structure and for its connecting structures (reactant(s)--transition state--product(s)) in order to evaluate relative energies.

Once launched for a high-T1 molecular structure, \textsc{autoCAS}
exploits a fast and non-converged complete-active-space (CAS) type calculation\cite{Stein2016} based
on the density matrix renormalization group (DMRG) 
algorithm \cite{White1992, White1993, Baiardi2020} to identify strongly correlated orbitals
for a CAS selection in a subsequent rigorous complete-active-space self-consistent field (CASSCF) calculation (based on standard CASSCF
exact diagonalization techniques or on DMRG)
with subsequent dynamic correlation treatment (typically by multi-reference perturbation theory).

Due to the huge number of valence electrons/orbitals to be inspected in the initial unconverged DMRG calculation, we exploited the large active space protocol described in Ref.~\citenum{Stein2019}.
We note that, even though the ground state was the only targeted state for all calculations in this work involving \textsc{autoCAS}, excited states can also be targeted\cite{Stein2017}.

We also note that one or more structures considered in these calculations for the extraction of multi-reference energy differences may turn out to be no multi-configurational cases at all, in which case an automated CAS selection will be difficult as there are no strongly correlated orbitals to select. Still, it can be ensured that a balanced active space can be selected over the whole reaction coordinate starting from those electronic structures that are most affected by static electron correlation.\cite{Stein2017}
For the HF and DMRG calculations required by \textsc{autoCAS} and the subsequent CASSCF\cite{Roos1980} and CASPT2\cite{Andersson1992} calculations, we employed the \textsc{ChronusQ} program package\cite{cqBeta, Li2019} in combination with the \textsc{QCMaquis} DMRG program\cite{Keller2015}.
The entire pipeline has been implemented as an AiiDA workflow \cite{aiida-autocas} in order to easily direct different steps onto different hardware: Steps such as DMRG and CASPT2 calculations require large-memory machines while other steps can be done on regular machines. Running the workflow over heterogeneous hardware is a key cost-saving factor on the cloud.

\subsection{Automation of all steps} \label{automation}
So far, we have introduced various tools to cover the different aspects of high-throughput first-principles reaction mechanism explorations. However, their number and their involved usage could create many challenges on the operator side (manual steering, error-prone manual data manipulation, \textit{etc.}). In order to alleviate these technical issues, all modules must be connected in an autonomous fashion. 
Therefore, one of the key components of the final AutoRXN workflow is to organize, automate, and streamline massive amounts of operations  in a way that is as autonomous as possible. The AutoRXN workflow can be divided into three main steps as shown in Figure \ref{fig:cloud_architecture}.

AiiDA is a well established workflow manager in computational chemistry and materials science to streamline complicated tasks into workflows.\cite{pizzi2016} However, after a careful examination of the computation characters of all steps, we decided to introduce a heterogeneous setting of workflow management: we activate the \textsc{Chemoton} meta program together with \textsc{Puffin} \cite{unsleber2022}
to manage the reaction exploration, but rely on AiiDA to steer coupled-cluster and multi-reference workflows.
We have compiled the meta problem described in the left of Figure~\ref{fig:cloud_architecture} into a container to include the mechanism exploration program \textsc{Chemoton}\cite{Moritz2022,unsleber2022}, the electronic structure model for fast tight-binding calculations \textsc{xTB}\cite{bannwarth2021}, the general open-source electronic
structure package \textsc{Serenity}\cite{Barton2021}, and all other dependencies for the reaction exploration. Note that not all of these components (in particular \textsc{xTB} and \textsc{Serenity}) were used in the current work. However, the presence of these additional components greatly enhances the versatility of the container for future applications. The benefit of the container is that it can easily orchestrate hundreds to thousands of cores for high-throughput execution hardware. This benefits especially from the Azure cloud, as VMs
with different micro-architectures can all be leveraged to extend the scale of exploration. 

Several tasks are then executed in this meta program in a specific order: 1) Initialization of the chemical systems, which includes general Cartesian structure set-up, conformer generation, and geometry optimization of any reactant or guiding species provided to \textsc{Chemoton} from an operator. 2) Elementary step trials include Newton trajectory calculations, transition state optimizations, internal reaction coordinate calculations to obtain minimal-energy reaction paths. 3) Sorting and clustering of the explored elementary steps and species into reactions and compounds. 4) Refinement of the structures along the minimal-energy paths using more accurate electronic structure calculations (such as coupled cluster), if necessary. 

Note that because in the reaction exploration (esp.~task 2), a massive amount of calculations can be initiated and evaluated, AiiDA as a daemon-based workflow manager will introduce a significant communication overhead to slow down the whole process. Therefore, we leave the job scheduling and management to a light-weight message queue, which allows \textsc{Chemoton} and \textsc{Puffin} communicate through \textsc{MongoDB} \cite{mongodb} inside the container of the meta program. The outcome of the meta program is a database that contains DFT optimized structures and reaction network information for further \textit{ab initio} calculations. 

The high-throughput coupled-cluster-calculations pipeline is a standalone AiiDA workflow\cite{aiida-nwchemex} and serves as the second step in the overall AutoRXN workflow. Unlike DFT or even lower-level calculations (\textit{e.g.}, the GFN2 tight-binding model\cite{bannwarth2019}) that can effciently be run on a single process or multi-processes on a single node, the CCSD(T) calculations leverage a more complex hardware architecture on runtime for significant acceleration. The communication between GPUs 
could be fragile and sometimes interrupted (\textit{e.g.}, GPU overheating could lead to a GPU communication issue). In addition, as we are using low-priority VMs 
in Azure, nodes could be evicted during busy hours which leads to a random job failure. For these reasons, the coupled-cluster workflow is executed as a standalone AiiDA workflow to ensure better cross-node parallelization, robust error handling, and a restart mechanism. We observed a $10\%$ job failure rate in the CCSD(T) calculations, which required a restart option. The data provenance and caching mechanism 
of AiiDA, under these circumstances, can restart the calculations from the last successful step, which saves a significant amount of time and costs for re-running failed jobs. 

The automated multi-reference-calculations pipeline is another standalone AiiDA workflow and serves as the final step in the overall catalysis workflow designed to take care of static-correlation cases, for which single-reference coupled cluster may exhibit problems. The benefit of using AiiDA here is similar to that for the coupled cluster workflow as various types of electronic structure calculations in the procedure require a reliable error handling and restart mechanisms. 

For the specific iron-complex catalyzed asymmetric hydrogenation reaction considered in this work, the three-step workflow was conducted in a semi-automated way, \textit{i.e.}, a manual invocation was required to trigger subtasks of exploration, coupled-cluster evaluation, and multi-reference validation. This was due to the dynamic and iterative nature of the workflow code development. However, the streamlined three steps can be automated in combination with database queries as a gigantic AiiDA workflow. A case study using the fully automated workflow will be demonstrated in forthcoming work.

\section{Results}\label{results}
After having described the conceptual and detailed computational settings for the AutoRXN workflow,
we now discuss the results that we obtained for the asymmetric hydrogen-transfer catalyst.

\subsection{Reaction network}\label{network}
The initial stage of the reaction network explorations yielded the minimal expected catalytic cycle for both \textit{(R)} and \textit{(S)} enantiomer of 1-phenylethanol.
According to the DFT calculations, the energetics for the two paths are not exactly the same. For the \textit{(S)} enantiomer, the barrier is 0.8\,kJ/mol and the reaction energy is $-$7.2\,kJ/mol, whereas for the \textit{(R)} enantiomer, the barrier is 0.9\,kJ/mol and the reaction energy is $-$15.7\,kJ/mol. The reason for this difference can be found in slightly different structures obtained: The stereoisomers of the products differ by 2\,kJ/mol, whereas this difference amounts to 6.5\,kJ/mol for the reactants; this is mostly due to a slightly different positioning of the two PH$_2$ ligands, which are closer to the substrate in the case of the less stable reactant stereoisomer.

\begin{figure}
    \centering
    \includegraphics[width=0.5\textwidth]{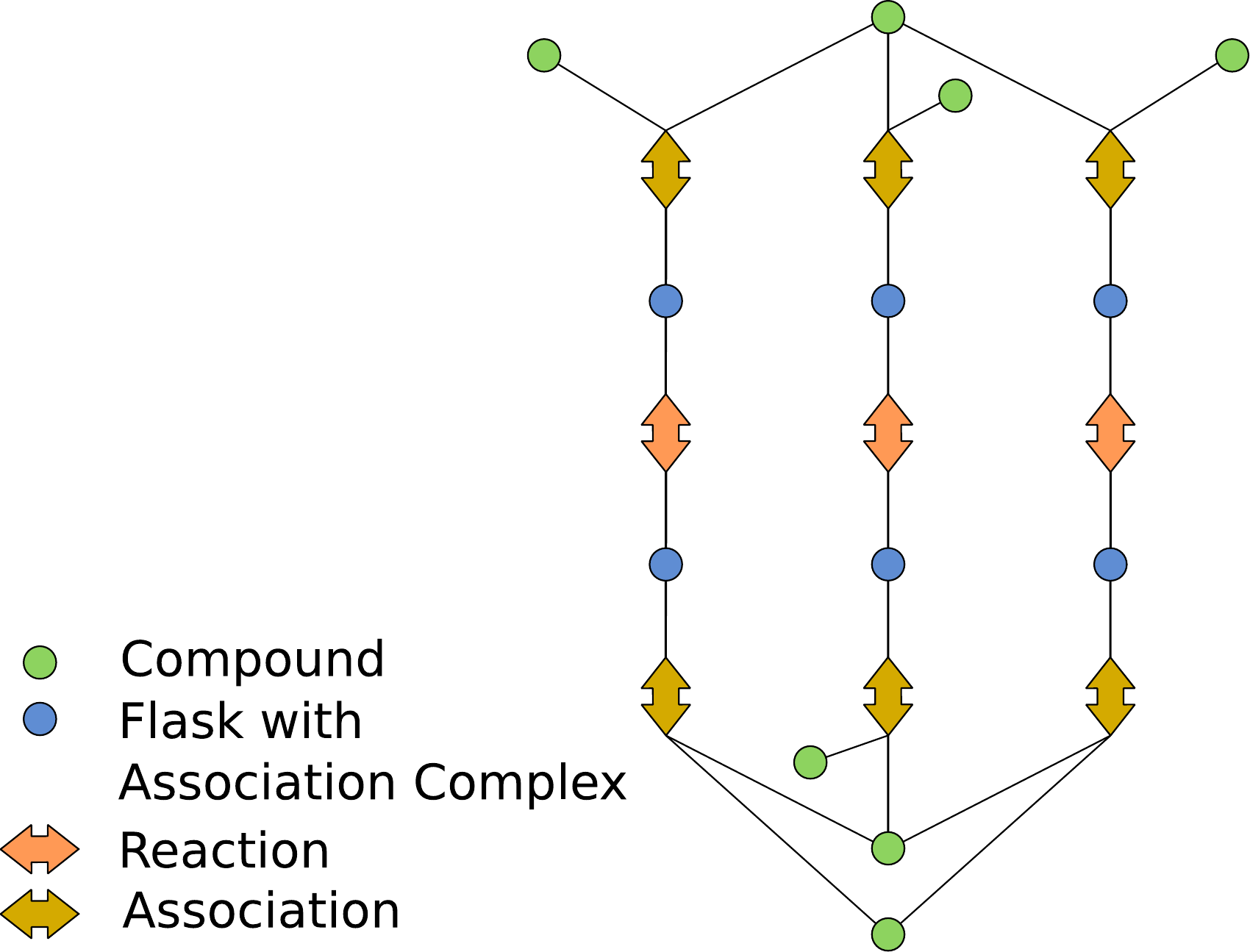}
    \caption{Catalytic cycles in a graph representation of the chemical reaction network shown in Fig.~\ref{fig:ath_cycle_lewis}.}
    \label{fig:network_cycle}
\end{figure}

The catalytic cycle shown in Figure~\ref{fig:ath_cycle_lewis} can be mapped onto a chemical reaction network graph; this graph is the result of the initial exploration phase.
This minimal view of the reaction network graph is shown in Figure~\ref{fig:network_cycle}.
The top center compound is the non-loaded catalyst; the chain of reactions down the center of the graph represent the steps required to load the catalyst with the substrate.
The term 'flask' is used as we have defined it in previous work\cite{Simm2017,unsleber2022}. We note that, in this particular case, only flasks of associated complexes of up to two molecules were set up.
The reaction chains on the left and right hand side represent the steps required to generate the \textit{(S)} and \textit{(R)} stereoisomers of the resulting product.
The view is arranged to fit the one of Figure~\ref{fig:ath_cycle_lewis}.
This small network represents those nodes/compounds that we would expect to be inspected by a manual exploration.

The AutoRXN workflow was then employed to generate further nodes in the reaction network.
Since we limited the run time to several months, the resulting network
features more details, but cannot be considered complete or exhaustively explored (as discussed above).
When the reaction network exploration was stopped, a total of 540 reactions and 2227 elementary steps were discovered.
Accordingly, the distribution of side reactions found across the initial cycle does not reflect an equal level of depth for all nodes of the minimal cycle. However, the results obtained already indicate
the depth that can and should be achieved in computational
campaigns that aim at the elucidation of a complete picture of the
reactivity of a catalyst.

\begin{figure}
    \centering
    \includegraphics[width=0.7\textwidth]{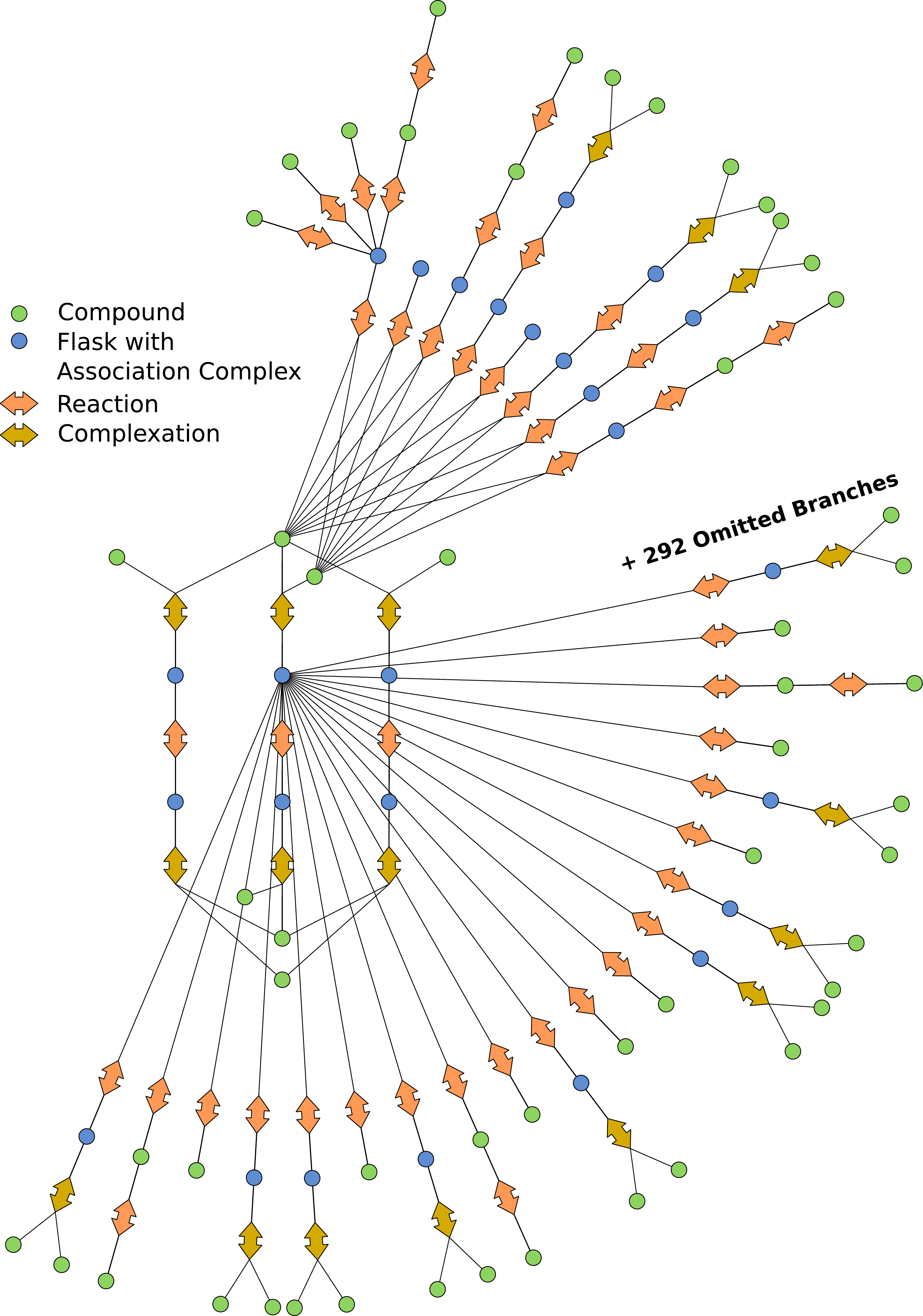}
    \caption{Network view indicating scope of the exploration. Note that 292 additional branches have been omitted from the representation for the sake of clarity.}
    \label{fig:network_complete}
\end{figure}

Figure~\ref{fig:network_complete} shows the branches that were explored beyond the minimal catalytic cycles.
These compounds are the catalyst (without a formal dihydrogen moiety loaded to it) and isopropanol, as well as the non-bonded complex of the catalyst and isopropanol.
Despite the fact that the imbalance of structures found is an artifact of the order in which the elementary step trials were set up and executed in our exploration campaign, 
the results obtained already allow us to draw some conclusions about alternative reaction paths for the initial step, which is the reason why we deemed this level of depth sufficient for our purposes here and did not further explore the mechanism to reduce the imbalance.

\begin{figure}
    \centering
    \includegraphics[width=15cm]{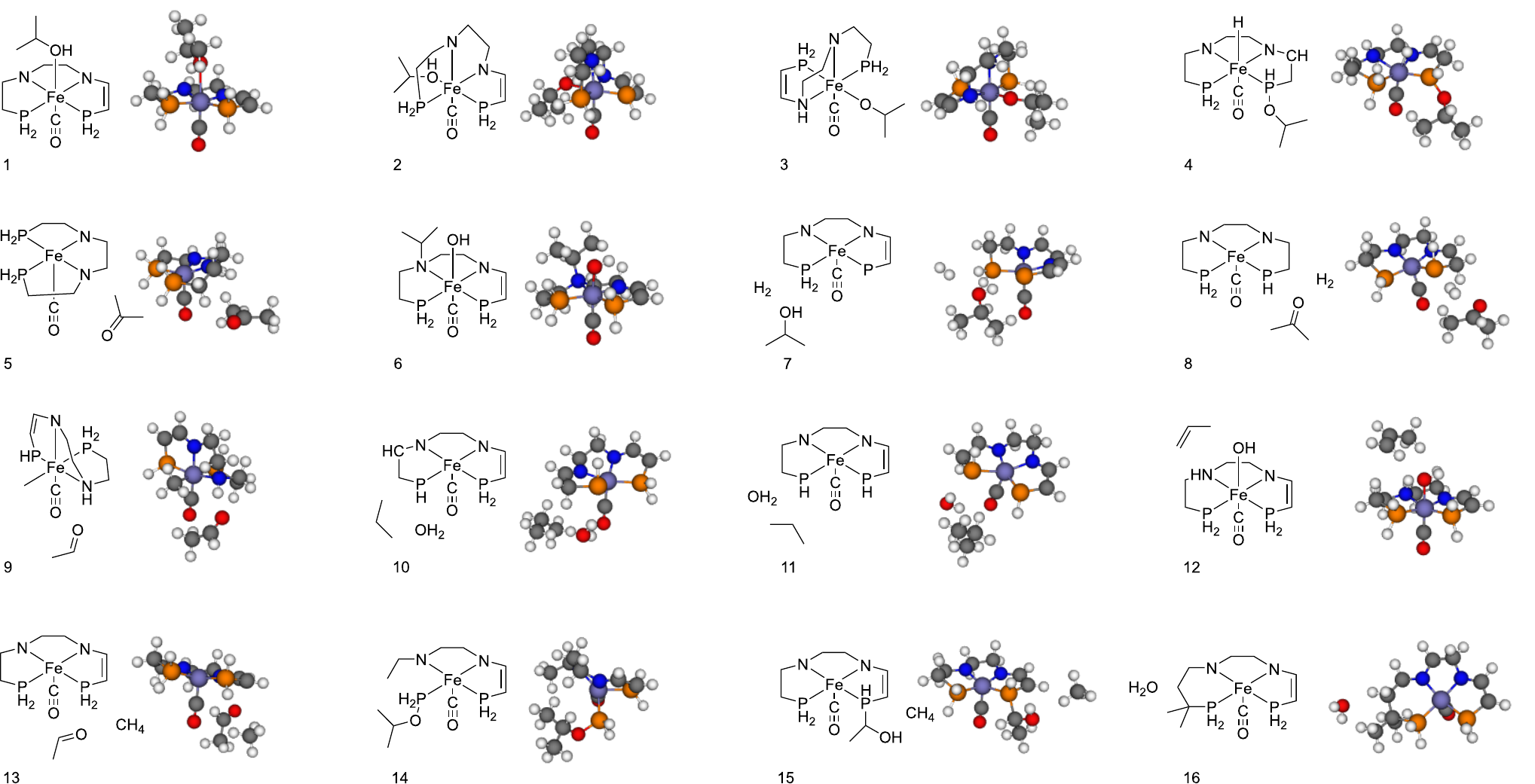}
    \caption{Products of a few selected side reactions found
    during the exploration.}
    \label{fig:structures}
\end{figure}

When analyzing the reactions identified during the exploration, we find that 23 reactions have a barrier of less than 100\,kJ/mol. These include all three reactions from the catalytic cycle. While the formal abstraction of two hydrogen atoms from isopropanol to load the catalyst (middle reaction path in Fig.~\ref{fig:ath_cycle_lewis}) is slightly endothermic with 22.6\,kJ/mol, the other two reactions of the catalytic cycle are exothermic.

\begin{figure}
    \centering
    \includegraphics[width=10cm]{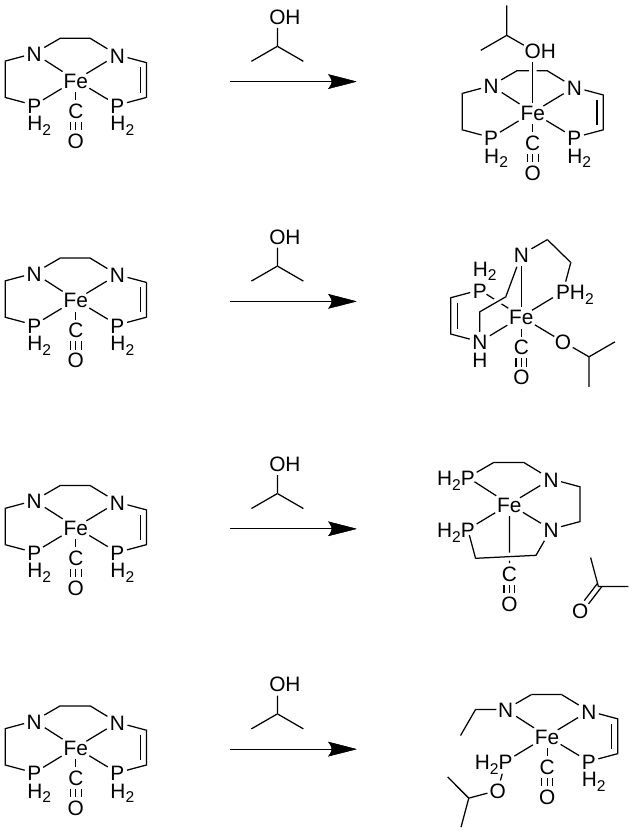}
    \caption{Representative examples of the four main classes of side reactions identified. From top to bottom: coordination of isopropanol to the catalyst; coordination of a solvent molecule to the catalyst with formal transfer of a hydrogen atom; (partial) hydrogenation of the C--C double bond; partial destruction of the chelate ligand.}
    \label{fig:reactions_lewis}
\end{figure}

The remaining 20 reactions with barriers below 100\,kJ/mol are side reactions of the (unloaded) catalyst and isopropanol. Only four of them are thermoneutral or slightly exothermic (in the range of $-$2\,kJ/mol to $-$16\,kJ/mol). Most reactions can be assigned to one of a few classes of similar reactions (\textit{cf.}, Fig.~\ref{fig:reactions_lewis}).
In a first class, isopropanol simply coordinates to the metal center of the catalyst. Depending on the direction
from which isopropanol coordinates to the iron atom, the complex is more or less distorted or rearranged (\textit{cf.}, 1 and 2 in Fig.~\ref{fig:structures}). In a second class of reactions, isopropanol also binds (or at least associates) to the catalyst, but in addition formally transfers one or two hydrogen atoms. These are preferentially transferred to the metal center, the nitrogen atoms, or the carbon atoms involved in the C=C double bond (\textit{cf.}, 3 and 4 in Fig.~\ref{fig:structures}). Finally, we can identify a third set of reactions in which the C=C double bond of the catalyst is (partially or fully) hydrogenated (\textit{e.g.}, 5 in Fig.~\ref{fig:structures}).

We also investigated all reactions with a barrier between 100\,kJ/mol and 200\,kJ/mol and a reaction energy of less than 60\,kJ/mol; we found 32 of such reactions in our exploration (again, these reactions involve only the catalyst and the solvent as reactants). Also here, most reactions belong to one of the three classes mentioned above (see Fig.~\ref{fig:reactions_lewis}).
However, we also find a few new reactions. For example, there are three reactions in which the hydroxy group of isopropanol binds to the metal center and the rest of this solvent molecule is transferred
to a nitrogen atom (\textit{cf.}, 6 in Fig.~\ref{fig:structures}). Furthermore, we find two reactions in which molecular hydrogen is produced (7 and 8 in Fig.~\ref{fig:structures}; in these cases, the solvent attacks the PH\textsubscript{2} group from the catalyst), and one in which acetaldehyde is created (and a methyl group from the solvent binds to the metal center, \textit{cf.}, 9 in Fig.~\ref{fig:structures}).

Analyzing the 33 reactions with barriers between 200\,kJ/mol and 300\,kJ/mol and reaction energies below 60\,kJ/mol, we find that there are almost no reactions in which isopropanol simply coordinates to the metal center. However, there are still many reactions belonging to the second class (hydrogen transfer to various parts of the catalyst) and the third class (attack of the C=C double bond). We also find six reactions in which acetaldehyde is produced and four in which molecular hydrogen is created (again by the solvent attacking the PH\textsubscript{2} group). Interestingly, there are also two reactions in which propane and water are produced from isopropanol by formally abstracting two hydrogen atoms from the catalyst (preferentially the PH\textsubscript{2} groups), \textit{cf.}, 10 and 11 in Fig.~\ref{fig:structures}. In another interesting reaction, propene is produced while the hydroxy group from the solvent is transferred to the metal center (12 in Fig.~\ref{fig:structures}). Finally, we also find two reactions in which methane is created (\textit{cf.}, 13 in Fig.~\ref{fig:structures}). Lastly, there is a new class of reactions (see Fig.~\ref{fig:reactions_lewis})
in which the C--P bond is broken, partially destroying the chelate ligand (\textit{e.g.}, 14 in Fig.~\ref{fig:structures}). The reactions in this class are particularly interesting, since these are slightly exothermic (reaction energies range from $-$4.8\,kJ/mol to $-$41.2\,kJ/mol); most other reactions are endothermic.

Finally, we analyzed all other exothermic reactions, irrespective of their barrier height, of which there are 13 (also here, the catalyst and isopropanol are the only reactants). With $-$61.8\,kJ/mol, the most exothermic reaction is one in which propane and water are produced from the solvent, formally abstracting two hydrogen atoms from the catalyst; 
there is also another elementary step of the same reaction with a reaction energy of $-$49.6\,kJ/mol. Again, we find a reaction in which isopropanol is inserted into the C--P bond (reaction energy $-$42\,kJ/mol). Most other reactions belong to the third class identified above (in which the C--C double bond is attacked). Also in this set of reactions, the PH\textsubscript{2} group is involved in many of the reactions. 
However, we find no reaction in which the PH\textsubscript{2} group is attacked by the solvent
and molecular hydrogen is produced. Instead, we find two reactions forming methane (15 in Fig.~\ref{fig:structures}) and a very interesting one in which a six-membered ring is created (16 in Fig.~\ref{fig:structures}).

\subsection{Coupled-cluster scrutinization of the reaction mechanism}\label{maincycle}

\begin{figure}
    \centering
    \includegraphics[width=16cm]{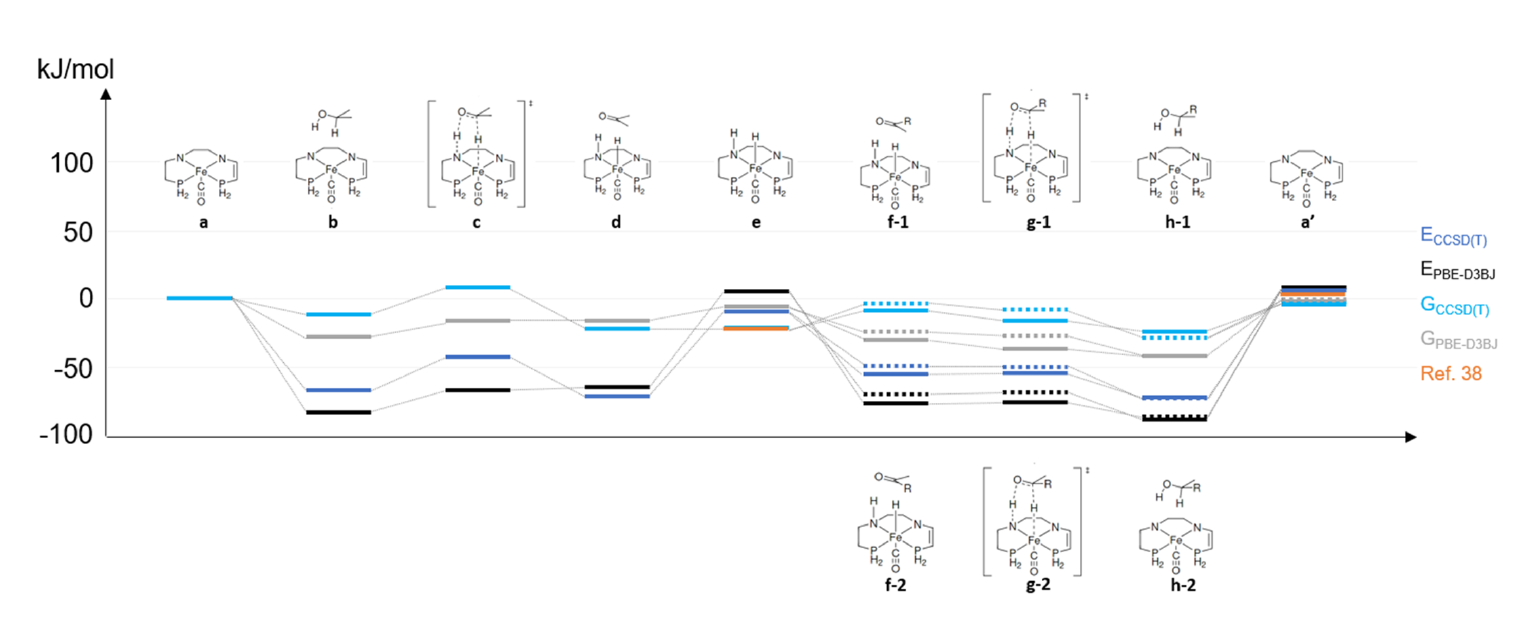}
    \caption{Energy diagram along the catalytic cycle of acetophenone hydrogenation. $a\xrightarrow{}e$ is the hydrogen loading step and $e\xrightarrow{}a'$ is the hydrogen transfer step.
    Dashed-line indicates asymmetric hydrogenation cycle for 1-phenylethanol enatiomers. The zero-energy reference is chosen to be the total energy of the iron catalyst, isopropanol, and acetophenone.}
    \label{fig:mainpathway}
\end{figure}

The main energy pathways of the asymmetric hydrogen-transfer reaction evaluated by PBE-D3BJ (black) and CCSD(T) (blue) are plotted in Fig.~\ref{fig:mainpathway}. We use $a-h$ to label the reactant, product, and intermediate configurations in the catalytic cycle. The catalytic cycle starts with isopropanol, acetophenone, and $a$, and ends with $a'$, acetone, and 1-phenylethanol. 
Though a step-wise hydrogen loading and transfer mechanism had been reported for the same simplified catalyst \cite{Prokopchuk2012}, more recent studies favor a concerted hydrogen loading and transfer mechanism. \cite{Zuo2013,Zuo2014,Zuo2016} Note, however, that in this work here,
we did not consider any solvation models (as this would open a new dimension of electronic structure modeling that is not important for the development of the AutoRXN workflow), charge-production reactions are intrinsically disfavored (as it would be the case in the gas phase). Ions cannot be stabilized without a dielectric environment and a step-wise mechanism is destabilized. Therefore, in our exploration, we focused on finding all relevant configurations for the concerted mechanism.

We first examine the electronic energies of the catalytic cycle. A general observation of the electronic energy diagram is that
PBE-D3BJ appears to predict lower reaction barriers compared with CCSD(T), 
especially in the elementary step of $b\xrightarrow{}c\xrightarrow{}d$, where the PBE-D3BJ electronic energy fails to show a typical 
reaction profile with some activation barrier.
PBE-D3BJ indicate that both hydrogen loading from isopropanol to the catalyst (+5.3\,kJ/mol) and hydrogenation to form 1-phenylethanol (+2.7\,kJ/mol ) are basically thermoneutral (\textit{i.e.}, the reaction energies are only very weakly endothermic). However, the CCSD(T) calculations show a slightly exothermic hydrogen loading step ($-$9.9\,kJ/mol), whereas hydrogenation to form 1-phenylethanol is endothermic (+15.6\,kJ/mol). However, also in this case these small energies can still be considered essentially thermoneutral.
In the electronic energy pathway, both PBE-D3BJ and CCSD(T) results suggest the acetone dissociation to be the rate-limiting step in the reaction process (+69.9\,kJ/mol with PBE-D3BJ, +60.26\,kJ/mol with CCSD(T)), whereas in all earlier studies, the rate-limiting step is the hydrogen loading. \cite{Prokopchuk2012,Zuo2014,Zuo2016}

When we now consider PBE-D3BJ Gibbs free-energy corrections (at 298\,K and 1\,atmosphere) to the electronic energies, the acetone dissociation energy has significantly dropped (+10.2\,kJ/mol for PBE-D3BJ and +0.58\,kJ/mol for CCSD(T)) the hydrogen loading has become the rate-limiting step of the reaction process (+11.2\,kJ/mol for PBE-D3BJ and +19.13\,kJ/mol for CCSD(T)). 

Both PBE-D3BJ+PBE-D3BJ (grey in Fig.~\ref{fig:mainpathway}) and CCSD(T)+PBE-D3BJ (cyan in Fig.~\ref{fig:mainpathway}) free energy pathways now give the same thermodynamic trends. 
The hydrogen loading step is exothermic (-6.2\,kJ/mol for PBE-D3BJ+PBE-D3BJ and -21.4\,kJ/mol for CCSD(T)+PBE-D3BJ) and the hydrogenation to form 1-phenylethanol is endothermic (+4.4\,kJ/mol for PBE-D3BJ+PBE-D3BJ and +17.42\,kJ/mol for CCSD(T)+PBE-D3BJ). 
At first sight, the CCSD(T) free-energy data agree with an earlier DFT study (orange in Fig.~ \ref{fig:mainpathway}, which corresponds to $-$21.7\,kJ/mol for hydrogen loading and +23.8\,kJ/mol for hydrogenation to form 1-phenylethanol)\cite{Prokopchuk2012} 
carried out with a different functional, different basis set, and including solvation effects.
This agreement could be coincidental because of the consideration of a solvent and the lack thereof. 

\begin{figure}
    \centering
    \includegraphics[width=16cm]{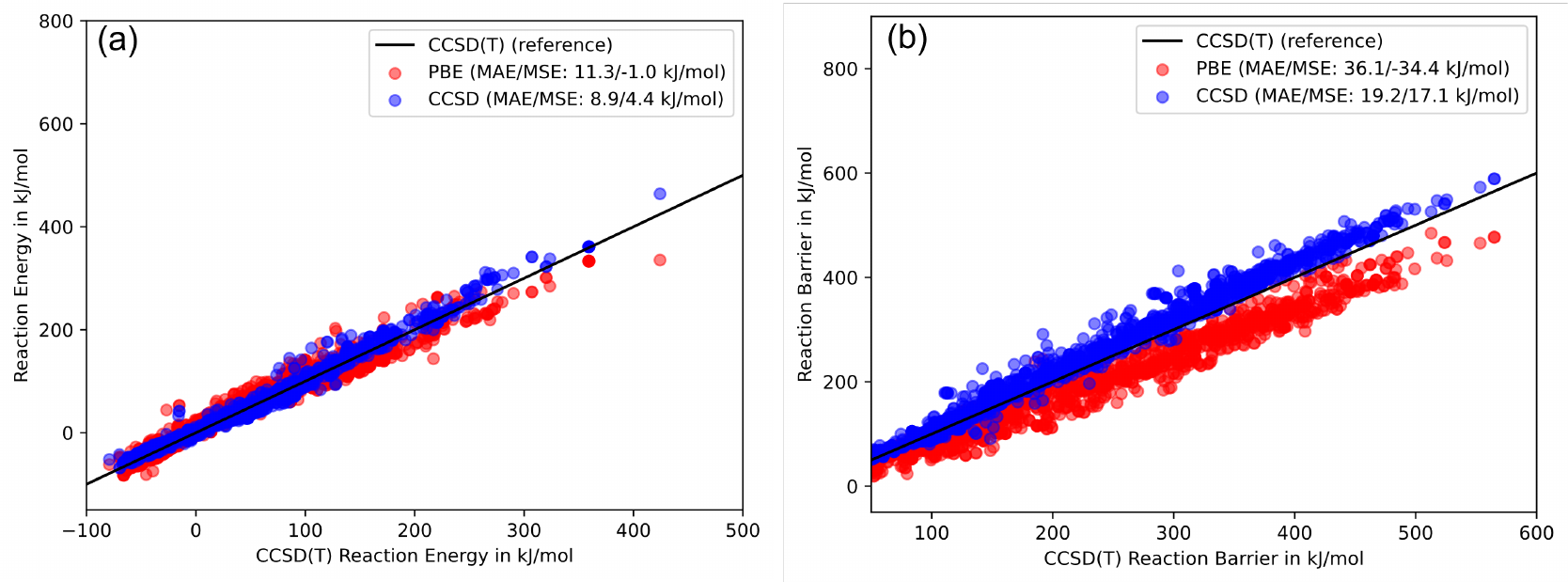}
    \caption{Correlation of a) reaction energies of all 2227 regular and barrierless elementary steps and b) reaction barriers of all 967 barrier-affected elementary steps calculated with PBE-D3BJ, and CCSD electronic structure models.
    CCSD(T) energies were taken as the reference and are shown as the solid back line.}
    \label{fig:stat}
\end{figure}

If we study the overall statistics of our DFT and CCSD(T) results across 2227 reaction energies and 967 reaction barriers, we find that for reaction energies, which are calculated between equilibrium structures, PBE-D3BJ results correlate very well with the CCSD(T) results with a mean absolute error close to 10\,kJ/mol and a bias (mean error)
of almost within 1\,kJ/mol. However, PBE-D3BJ systematically underestimates reaction barriers with a mean absolute error of 36.1\,kJ/mol and a bias of $-$34.4\,kJ/mol (Fig.~\ref{fig:stat}).

\subsection{Back checking multi-reference cases}\label{backchecking}
Since CCSD(T) is a single-reference method, its results may be compromised in multi-configurational cases.
The T1 diagnostics has become a well-established single-reference diagnostics for multi-reference characters.\cite{Lee1989}
In this work, we applied a relative loose T1 diagnostics threshold (0.1) to determine how many structures in the network exhibit a multi-reference character.
For coupled-cluster results with a T1 diagnostics larger than 0.1, the AutoRXN workflow subjected the structure to the multi-reference add-on pipeline in order to validate whether a significant multi-reference character can also be confirmed in a multi-configurational calculation and, if this test is positive, to produce a substitute result based on multi-reference perturbation theory. If the single-orbital-entropy\cite{legeza2004} based multi-reference diagnostic $Z_{s(1)}$ \cite{Stein2017b}
is larger than 0.14, a multi-reference character is suggested. The workflow will continue to conduct the active space selection and complete CASPT2 calculations fully automatically. This pipeline will be applied to all structures connected in the elementary step under consideration for a consistent set of activation energy and reaction energy at CASPT2 level. Note that a mixture of single-reference and multi-reference cases in one elementary step requires the activation of an automated orbital matching procedure to ensure balanced active space calculations, which we have presented in Ref.~\citenum{bensberg2022}.
 
\begin{figure}
    \centering
    \includegraphics[width=16cm]{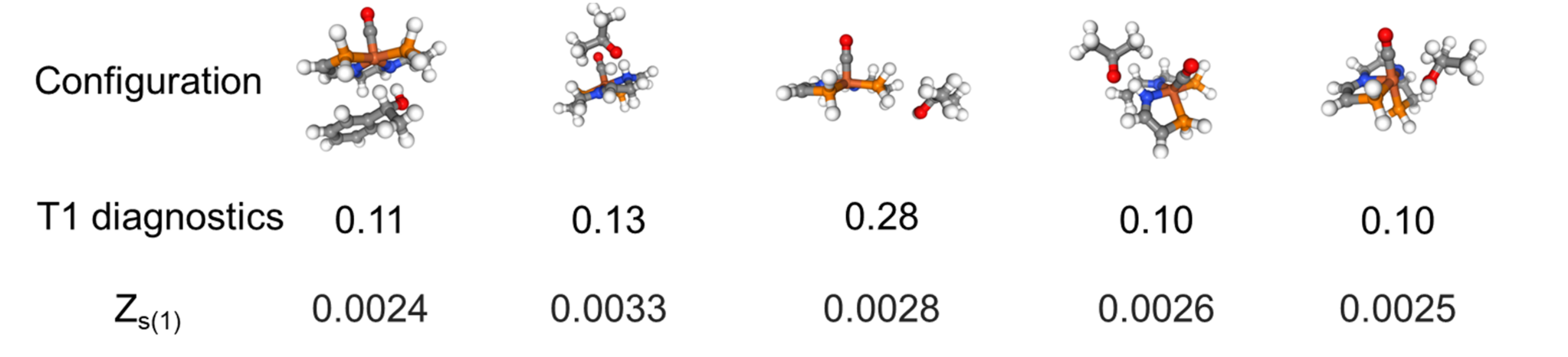}
    \caption{Three-dimensional molecular structures, T1 diagnostics, and orbital-entropy-based multi-reference diagnostics $Z_{s(1)}$ of high-T1 molecules.}
    \label{fig:autocas-t1}
\end{figure}

The structures and diagnostics of all high-T1 molecules in our network are listed in Fig.~\ref{fig:autocas-t1}.
Interestingly, all $Z_{s(1)}$ data for the critical structures (also
given in Fig.~\ref{fig:autocas-t1}) turn out to be very small and
hence provide no indication of actual multi-configurational cases.
Hence, we find the T1 diagnostics (which rest on single-reference data
and point to insufficiencies in the Hartree--Fock reference) to be
not reliable. Our multi-configurational measures point towards clear single-reference calculations, so that no further calculations are required for our specific example. Still, we performed such calculations, which confirmed the single-reference nature of all 
structures in Fig.~\ref{fig:autocas-t1}.

\section{Conclusions}\label{conclusion}
In this work, we established the computational AutoRXN workflow for
cloud-based high-throughput virtual screening of molecular structures, chemical mechanisms,
and reaction networks. At the example of a homogeneous catalyst
we explored a catalytic cycle at increased structural resolution,
highlighting paths for the gas phase that have not been considered for this catalyst so far. 
The AutoRXN workflow automatically subjected all optimized
stationary structures, \textit{i.e.}, stable intermediates and transition
state structures, to coupled-cluster single-point calculations as a means to
validate the DFT results. Although such a system-focused and built-in 
benchmarking\cite{Reiher2021} may be restricted to 
chemically differing molecular structures
according to a measure for molecular similarity\cite{Simm2018},
it has been the explicit purpose of this work to demonstrate that even
a complete set of coupled-cluster results can be obtained through
cloud-based high-throughput calculations. Moreover, we showed that
the single-reference coupled cluster results can be scrutinized
by automatically detecting multi-reference cases. These cases can then
be subjected to multi-configurational calculations. It is important
to emphasize that even these multi-configurational calculations
can be launched automatically through the \textsc{autoCAS} algorithm\cite{Stein2016,Stein2019},
for this work implemented as a command line interface that seamlessly interacts with
all components of the AutoRXN workflow.

We expect that the AutoRXN workflow will be beneficial for any type of
exploration into chemical reaction space: it will make first-principles explorations of
reaction mechanisms at unprecedented depth straightforwardly 
accessible with rather little pre-knowledge about the computational
procedures. Naturally, extension of the workflow, for instance, towards automated 
property calculations is easily possible. 

\section*{Author Declaration}
H.L., L.T., A.G., D.W., and M.T. declare a conflict of interest 
as the cloud service on which AutoRXN has been developed is the Microsoft Azure service.

\section*{Data Availability Statement}
The data that support the findings of this study are available within the article and in the Zenodo references cited in the main text.

\section*{Acknowledgement}
M.R. gratefully acknowledges financial support by the Swiss National Science Foundation (through Project No. 200021\_182400 and as part of the NCCR Catalysis, a National Centre of Competence in Research) and by ETH Zurich (Project No. ETH-43 20-2). A.P., B.P., and K.K. acknowledge the support from the Center for Scalable, Predictive methods for Excitation and Correlated phenomena (SPEC), which is funded by the U.S. Department of Energy (DOE), Office of Science, Office of Basic Energy Sciences, the Division of Chemical Sciences, Geosciences, and Biosciences. SPEC is located at Pacific Northwest National Laboratory (PNNL) operated for the U.S. Department of Energy by the Battelle Memorial Institute under Contract DE-AC06-76RLO-1830.

\providecommand{\latin}[1]{#1}
\makeatletter
\providecommand{\doi}
  {\begingroup\let\do\@makeother\dospecials
  \catcode`\{=1 \catcode`\}=2 \doi@aux}
\providecommand{\doi@aux}[1]{\endgroup\texttt{#1}}
\makeatother
\providecommand*\mcitethebibliography{\thebibliography}
\csname @ifundefined\endcsname{endmcitethebibliography}
  {\let\endmcitethebibliography\endthebibliography}{}

\end{document}